\documentclass{article}
\usepackage{authblk}
\usepackage{amsmath}
\usepackage{enumerate}
\usepackage[numbers]{natbib}
\usepackage{graphicx}
\IfFileExists{upquote.sty}{\usepackage{upquote}}{}

\begin{document}

\title{Multiplicity for a Group Sequential Trial with Biomarker Subpopulations}
\author[1]{Ting-Yu Chen}
\author[2]{Jing Zhao}
\author[2]{Linda Sun}
\author[2]{Keaven M. Anderson}
\affil[1]{The University of Texas Health Science Center at Houston, TX, USA}
\affil[2]{Merck \& Co., Inc., Kenilworth, NJ, USA}

\maketitle


\begin{abstract}
Biomarker subpopulations have become increasingly important for drug development in targeted therapies.  The use of biomarkers has the potential to facilitate more effective outcomes by guiding patient selection appropriately, thus enhancing the benefit-risk profile and improving trial power.  Studying a broad population simultaneously with a more targeted one allows the trial to determine the population for which a treatment is effective and allows a goal of making approved regulatory labeling as inclusive as is appropriate. We examine new methods accounting for the complete correlation structure in group sequential designs with hypotheses in nested subgroups. The designs provide full control of family-wise Type I error rate. This extension of previous methods accounting for either group sequential design or correlation between subgroups improves efficiency (power or sample size) over a typical Bonferroni approach for testing nested populations.
\end{abstract}

\section{Introduction}
Conventional design with only one primary study population (an overall population) has recently been challenged \cite{buysebiomarker, mandrekarbiomarker}, particularly when the treatment effect may be heterogeneous due to observable clinical or biologic/genomic characteristics.  For example, in recent oncology clinical trials, biomarker subpopulations (biomarker +/-) have become increasingly important for drug development in tailored therapies to fulfill regulatory commitments \cite{Wang2007}.  The use of biomarkers has the potential to facilitate the availability of safer and more effective drug or biotechnology products, to guide dose selection, and to enhance their benefit-risk profile \cite{FDA2014}.  While the overall population targets the goal of making the approved regulatory labeling as inclusive as possible, evaluating the benefit in a biomarker positive subpopulation can mitigate the risk that biomarker negative patients could dilute the efficacy in the overall population. 

In the situation of testing hypotheses in multiple populations, multiplicity needs to be carefully considered to ensure strong control of the family-wise Type I error (or family-wise error rate, FWER) and to maximize the study power.  In the setting described above, there is a known correlation structure in the asymptotic distribution for the joint test statistics across interim and final analysis as well as across populations.  To-date, people have designed trials accounting for correlation in interim timing \cite{JTBook} or subpopulations \cite{Spiessens2010, Holmgren2017}, respectively, but have not taken advantage of the full correlation structure including both populations and interim analyses, leading to stricter bounds than necessary to control Type I error.  Therefore, in this paper, we extend the group sequential design setup with multiple biomarker populations and develop the method and calculations with less conservative bounds, which results in a smaller required sample size or greater power, while controlling the FWER. This is a realization of the improved weighted parametric test mentioned in \cite{MaurerBretz2013} for multiple testing in group sequential trials using graphical approaches.

In section 2, we briefly review methods in group sequential design.  We introduce the complete correlation structure (CCS) incorporating both interim analyses and populations in Section 3.  In the same section, the calculations for adjusted nominal alpha levels,  power for hypothesis testing in each population and sample size are also presented.  The effect of CCS on clinical trial design is demonstrated in Section 4, and an application of CCS is shown in Section 5.  Discussion and extensions of CCS appear in Section 6. Example R program code is included in Appendix C and Appendix D. 

\section{Background}

Group sequential design has played an increasingly important role in modern clinical trials for ethical reasons and economic considerations.  In group sequential design, interim analyses are performed during the trial.  Further study follow-up may be stopped in accordance with a pre-defined stopping rule as soon as conclusive results are observed.  Therefore, a conclusion may be reached at an earlier stage with lower financial and human cost.  More importantly group sequential design offers the possibility to accelerate replacement of an inferior therapy by a superior one compared to fixed sample size study design where a decision only can be made at the end of the trial.  
One aspect of group sequential design is to control temporal correlation among interim analyses and the final analysis.  Pocock \cite{PocockBound, Pocock1982} and O’Brien and Fleming \cite{OF} initially popularized group sequential test procedures to manage multiplicity.

Lan and DeMets \cite{LanDeMets} introduced non-decreasing alpha-spending functions to determine interim efficacy bounds.  Their method loosened the rules required by the Pocock and O'Brien and Fleming approaches, allowing flexibility in timing of analysis.  They developed spending functions to approximate the Pocock and O’Brien and Fleming designs.  Kim and DeMets \cite{KimDeMets} as well as Hwang, Shih and DeCani \cite{HwangShihDeCani} also proposed flexible one-parameter families that again can approximate Pocock or O’Brien and Fleming as well as other boundaries.  
A comprehensive illustration of group sequential design can be found, for example, in Jennison and Turnbull \cite{JTBook}, Proschan, Lan, and Wittes \cite{PLWBook}, or Wassmer and Brannath \cite{WBBook}. Anderson and Clark \cite{AndClark} suggested 2-parameter spending functions that could be used to further customize bounds in a fit-for-purpose manner. 

Above we introduced methods that were applied to a single endpoint and population for a single hypothesis.  
However, it is common to have multiple hypotheses regarding different endpoints and populations with group sequential analysis, increasing the complexity of Type I error control.  
Here we consider the case of testing the treatment effect in nested populations within a single clinical trial.  Maurer and Bretz \cite{MaurerBretz2013} showed the usage of a graphical approach in group sequential design when multiple hypotheses were tested in a trial.  Later in Maurer, Glimm, and Bretz’s work \cite{Maurer2011}, they extended the algorithm of the test procedure by implementing group sequential boundaries and suggested its application to a comparison of multiple endpoints for a subgroup and an overall population. However, they did not provide detailed instruction on results for population correlations.   

Another useful application of group sequential method is to account for correlation among subgroups (sub-populations) and the overall population.  Spiessens and Debois \cite{Spiessens2010} suggested the correlation of test statistics between nested subgroup and the overall population can be addressed using the same method in interim analysis (i.e. group sequential design) because it could improve the efficiency of clinical trials while controlling the FWER.  Holmgren \cite{Holmgren2017} proposed a similar concept using group sequential design boundaries at early phase (e.g. phase II) in the decision of choosing a biomarker expression level used in later phase (phase III) trials.  However, none of these papers accounts for the benefit gained when simultaneously accounting for temporal and population correlation, a situation that arises increasingly in oncology clinical trials.

In this paper, we synthesize the concepts regarding population correlation from\cite{Spiessens2010, Holmgren2017} and incorporate it with \cite{MaurerBretz2013}  on multiplicity control for multiple hypotheses in group sequential design.  We propose a complete correlation structure (CCS) for the covariance matrix of test statistics that accounts for both temporal and population correlations in group sequential design to improve design efficiency.

\section{Methods}

\subsection{Complete Correlation Structure}

Jennison and Turnbull \cite{JTBook} summarize the asymptotic distribution of test statistics in group sequential design.  We extend this to the above circumstance by the proposed CCS method accounting for correlations among test statistics at multiple analyses over time (temporal perspective) as well as of multiple populations.  
We consider a 2-arm clinical trial comparing the treatment effect of an experimental treatment group to a control group in $I$  nested biomarker subgroups, including the overall population. Each population has an hypothesis,  and these hypotheses are evaluated at $K$ analyses.  We let $i$ denote the index for increasing nested populations, $i=1,\ldots,I,$ while $k$ represents the index for the stage of interim analyses and final analysis, $k=1,\ldots,K.$
Let $n_{ik}$ be the number of observations (or number of events for time-to-event endpoints) collected cumulatively through stage $k$ in population $i$.

Let $\theta_i$  represent the underlying effect for experimental vs. control treatment in population $i$ for $i=1,2,\ldots,I.$  We assume $H_{i0}$: $\theta_i=0$ represents no treatment effect for the experimental treatment relative to control in population $i$, while $H_{ia}$: $\theta_i=\theta_{ia}>0$ represents an advantage for experimental treatment. 
Let $Z_{ik}$ be the standardized test statistic for nested population $i$ at stage $k=1,\ldots,K,$ $i=1,…,I.$  We let $n_{ik}$ represent the sample size for a binomial or normal endpoint or the number of events for a treatment comparison in population $i$ at analysis $k.$ For each population, we assume $E(Z_{ik})=\theta_i \sqrt{n_{ik}},$ $Var(Z_{ik})=1;$ this is the so-called canonical form of \cite{JTBook}.  We will demonstrate that for cases where group sequential theory can be applied either across $k=1,\ldots,K$ or populations $i=1,\ldots,I$ that we further have a $K\times I$-variate normal distribution with

\begin{equation} \label{eq1}
\begin{split}
&\hbox{E}(Z_{ik})=\sqrt{n_{ik}}\theta_i\\
&\hbox{Var}(Z_{ik})=1
\end{split}
\end{equation}
with a known covariance structure for each pair $(Z_{ik}$, $Z_{i^\prime k^\prime})$  for $1\le i,i^\prime \le I,$ $1\le k,k^\prime\le K$  given by

\begin{equation}
\hbox{Cov}(Z_{ik},Z_{i^\prime k^\prime})
=\frac{n_{i\wedge i^\prime k\wedge k^\prime}}{\sqrt{n_{ik}n_{i^\prime k^\prime}}}
\end{equation}
where the operator $\wedge$ represents the minimum. In the numerator for this covariance for a time-to-event outcome, we have the number of events included in both (intersection) test statistics for which we are computing the covariance.  The denominator has the geometric mean of the events considered in each test statistic separately. For a normal or binary outcome, we would count observations rather than events. For $k=k^\prime,$ this is the result from \cite{Spiessens2010, Holmgren2017}.  For $i=i^\prime,$ this is a standard group sequential design result.  Combining these two results (e.g., independent increment in population and then an increment or decrement in time) yields the general result above.  With the entire correlation structure, we can calculate adjusted nominal alpha levels, population power, and sample size that produce a more efficient design than one that does not account for the entire correlation structure.  A detailed derivation is provided in Appendix A.

\subsection{Weighted Parametric Testing Bounds}

When working with sets of hypotheses, we will let $I$ represent the set of population indices rather than the number of populations $I$ as in Section 3.1. For a set of indices $J\subseteq I,$ we denote the intersection null hypothesis as

\begin{equation}
H_J=\cap_{i\in J}  H_{i0}. 
\end{equation}
The closed testing principle \cite{Marcus1976} means we can reject an individual null hypothesis $H_{i0}$ for population $i \in I$  if for all $J\subseteq I$ with $i\in J$ we can reject $H_J$.

We define a test of $H_J$ using a set of bounds $b_{ik}, i\in J, 1\le k\le K$.
With these bounds, we reject $H_J$ if for any $i \in J, 1\le k\le K$ we have $Z_{ik}\ge b_{ik}$.
As noted by \cite{MaurerBretz2013}, the usual sequentially rejective approach has not been proven acceptable for use here since we have not demonstrated consonance. To reject an individual null hypothesis $H_{i0}$  we must reject $H_J$ at level $\alpha$ for every $J\subseteq I$ with $i\in J.$
The complete correlation structure and multivariate normal distribution to can be used to compute the probability to reject $H_J$ under arbitrary values of $\theta_i$ for $i\in J$:

\begin{equation}
1- Pr\left(\cap_{i\in J, 1\le k\le K}\{Z_{ik}<b_{ik}\}|\theta_i, i\in J\right).
\end{equation}
We have used the {\bf mvtnorm} R package \cite{mvtnorm} for this computation.

\subsubsection{Computing Testing Bounds}

To set the above bounds to control Type I error for testing under the intersection hypothesis $H_J$ at level $\alpha$, we follow the allocation and reallocation approach of \cite{Bretz2011}.
Thus, we set a weights $w_i(I),$ for $i\in I$  and a transition matrix $G = (g_{ij}), i,j\in I$.
For subset $J\subseteq I$,  these weights can be translated into weights $w_i(J)$ for $i\in J$ to test $H_J$.
The information fraction at stage $k$ for population $i$ is $t_{ik}= n_{ik}/n_{iK}.$ 
While we generally assume $t_{ik}=t_k$ across populations, this is not necessary. 
We define a spending function family as in \cite{MaurerBretz2013} for each population as $f_i(t;\gamma)$ for $0<\gamma<1$ and $t\ge 0$.
We further require $f_i(0;\gamma)=0$, $f_i(t;\gamma)=1$ for $t\ge 1$, $f_i(t;\gamma)$ increasing in both $t$ and $\gamma$ and the other conditions on spending function families of \cite{MaurerBretz2013}.

We present 3 algorthims for deriving testing bounds for $H_J, J\subseteq I.$ 
This involves both deriving bounds at the time of design using planned information (sample size or event counts at each analysis), as well as updating bounds if observed information at an analysis (interim or final) differs from planned. The first algorithm is presented here and the other two in Appendix B.
The 3 approaches can be described briefly as follows:

\begin{description}
\item[Algorithm 1] Adjust current and future bounds at time of each analysis.
\item[Algorithm 2] Adjust current bounds at time of each analysis.
\item[Algorithm 3] Allocate excess $\alpha$ to largest population only. 
\end{description}

{\noindent \bf Algorithm 1}

For a given $J\subseteq I$ and sequentially for analysis $k\in 1,\ldots,K$:

\begin{enumerate}
\item Assume for $j<k$ that bounds $b_{ij}=c_{ij}(J), j<k, i\in J$ have already been set. These will remain unchanged once finalized at analysis $k$.
\item Based on updated $n_{ik}$ and future planned $n_{ij}, k<j\le K, i\in J$ bounds for analyses $j=k,\ldots K$ are re-planned.
\begin{enumerate}[a.] 
\item Choose a nominal $\alpha_k^*(J)\ge \alpha$.
\item Set $b_{ij}$ for $k\le j\le K$ to control Type I error for hypothesis $H_{i0}$ at level $f_i(t_{ij};\alpha_k^*(J)\times w_i(J))$ at analysis $j$; {\it i.e.,} 
\begin{equation}
\begin{split}
1-Pr&(\{Z_{ij}< b_{ij}\}\cap_{1\le j^\prime<j}\{Z_{ij^\prime}<c_{ij^\prime}(J)\} |H_{i0}) \\
&=f_i(t_{ij}; w_i(J)\times\alpha_k^*(J)).
\end{split}
\end{equation}
This can be done with available group sequential design software such as in the {\bf gsDesign} R package \cite{gsDesign}.
\item  Update $\alpha_k^*(J)$ until the above bounds control the overall testing level for $H_J$ at level $\alpha$; {\it i.e.,}
\begin{equation}
1 - Pr(\bigcap_{i\in J}\left\{\cap_{1\le j^<k}\{Z_{ij}<c_{ij}(J)\}
       \cap_{k\le j\le K}\{Z_{ij}<b_{ij}\}\right\}|H_J) = \alpha.
\end{equation}
This can be done as noted in equation (4).
\item After the appropriate $\alpha_k^*(J)$ has been derived, we set $c_{ik}(J)=b_{ik}$.
\end{enumerate}
\end{enumerate}

\section{Evaluating design properties}

\subsection{Population Power and Sample Size}

When a study is designed, the CCS can be calculated by prevalence of biomarker populations and the design interim analysis timing.
Here we assume the information accumulates at the same rate for different populations under the null and alternative hypothesis to simplify. 
For example, consider a study with 2 populations (biomarker subgroup and overall population) and 3 analyses (two interim analyses and one final analysis). 
Assume further that the prevalence of the biomarker subgroup is $p,$ and the design interim analysis timing based on information fraction is $t_1$ and $t_2$ at interims 1 and 2, respectively.
The covariance matrix for test statistics 
$(Z_{11},Z_{21},Z_{12},Z_{22},Z_{13},Z_{23})$ is as same as in Section 3.3 and can be calculated as  

\begin{equation}
\begin{bmatrix}
1 & \sqrt p & \sqrt{t_1/t_2} & \sqrt{pt_1/t_2} &\sqrt{t_1} &\sqrt{pt_1}\\
  & 1 &\sqrt{pt_1/t_2}  &\sqrt{t_1/t_2} &\sqrt{pt_1} &\sqrt{t_1}\\
  &   & 1 &\sqrt p &\sqrt{t_2} &\sqrt{pt_2}\\
  &   &   &1 &\sqrt{pt_2} &\sqrt{t_2}\\
  &   &   &  &1 &\sqrt p\\
  &   &  &  &  &1\\

\end{bmatrix}
\end{equation}

With the CCS, we find $c_{ij}(J)$ as in Algorithm 1 above, providing more generous bounds than a Bonferroni-adjustment approach.
This leads to either a smaller sample size requirement or greater power as demonstrated below in Section 5.

\subsection{Effect of Biomarker Prevalence}

Next, we demonstrate the impact of incorporating CCS in a simple case with only one interim analysis ({\it i.e.,} $K=2$) and one subpopulation ({\it i.e.,} $I=2$) planned; example R code is provided in Appendix C.  
Assuming the FWER is equally split to the subgroup and the overall population (i.e. $w_1=0.5, w_2=0.5$).  We use the Lan and DeMets spending function approximating O’Brien and Fleming bounds.  We also assume the interim analysis information fraction ($t_{i1}$; proportion of final planned events) at 0.5 and adjust subgroup proportion from 0.3 to 0.8 to investigate effect caused by the prevalence of the biomarker subgroup. 
Define, $p_{ik}=n_{ik}/n_{Ik},$ which is the same for $k = 1, 2,$ if the information accumulates at the same rate for the two populations. Additionally, the effect size for subgroup is set at 0.15.  
In order to maintain the same power for both populations, we set the effect size for overall population with a proportion of $\sqrt p$ the effect size in subgroup (i.e. effect size 0.106 for overall population when the prevalence of biomarker subgroup is p = 0.5).

Figures 1-3 showed the effect of biomarker prevalence ($p_{1k}$) to adjusted nominal $\alpha$-level ($w_i (I)\alpha_K^* (I)$), population power, and sample size.

\begin{figure}
\centering
\includegraphics{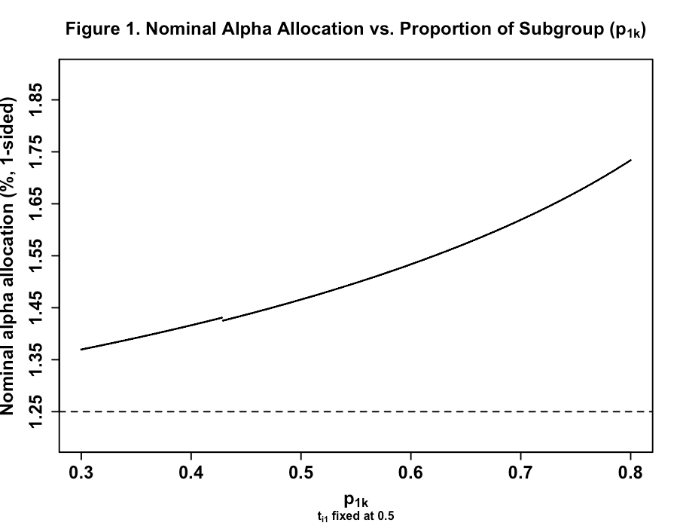}
\end{figure}

\begin{figure}
\centering
\includegraphics{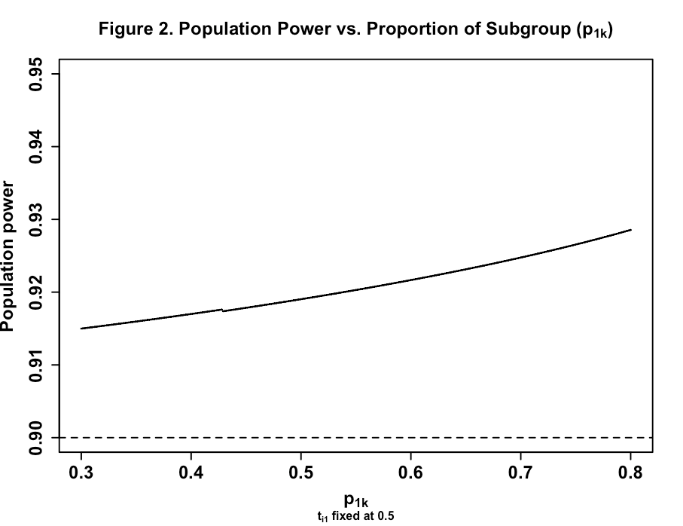}
\end{figure}

\begin{figure}
\centering
\includegraphics{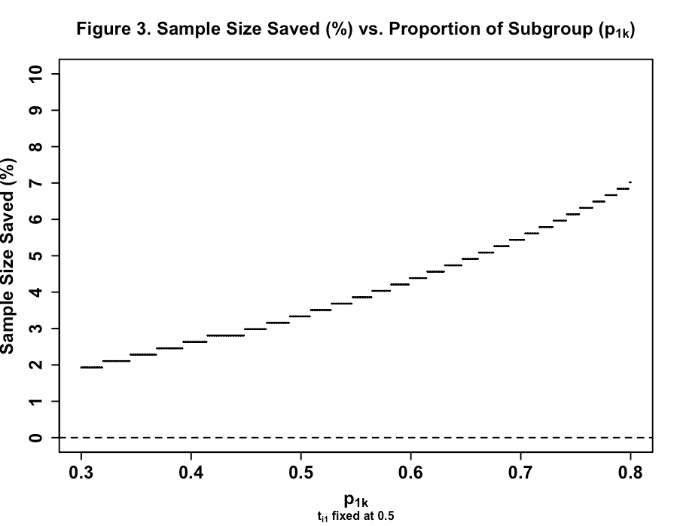}
\end{figure}

\section{Example in an Oncology Clinical Trial}

In this section, we show the implementation of CCS in a hypothetical oncology clinical trial.  Assume a 2-arm trial with a primary endpoint of overall survival (OS).
Under the complete null hypothesis the distribution of time-to-death is assumed to follow an exponential distribution with a median of 17.5 months in both the treatment groups and populations.
The drop-out rate is assumed to be 3\% annually.  
The FWER of 2.5\% (1-sided) is equally distributed to the subgroup and all subjects.
For the alternate hypothesis, we assume the hazard ratio (HR) for OS is 0.65 in the subgroup and HR=0.7 in the overall population.  
We expect 434 OS events overall with 296 in subjects with biomarker (+) with 90\% power in both populations.   The first interim analysis (IA1) is planned at 50\% of final planned events, the second interim analysis (IA2) is planned at 75\% of final planned, and the final analysis (FA) is conducted when all targeted event counts have been achieved.  
The prevalence of biomarker (+) subgroup is assumed to be 60\%.  For detailed R program code, please see Appendix D.  

As part of the above, we used the CCS and Algorithm 1 to find $c_{ij}(J)$. 
We used equation (4) to compute the power for these bounds and adjust the sample size to achieve 90\% power for each population at the adjusted nominal $\alpha$-level generated using Algorithm 1 under the complete null hypothesis.

Table 1 below shows the comparison of Bonferroni-adjusted group sequential design (adjusted with temporal correlations only) and CCS group sequential design (adjusted with both temporal and population correlations) in several elements. The nominal testing level of 1.53\% vs. 1.25\% using Bonferroni may seem small; however, we note that the sum of the 2 nominal alpha levels for group sequential testing is 1.277\% (vs 1.25\% which is Bonferroni approach for interim analysis and final analysis without considering temporal correlation) when testing half-way through the trial with overall alpha=1.25\% and O’Brien-Fleming-like spending. Thus, the gains from incorporating population correlations here are greater than those correlations over time. This is largely due to the substantial spend on each population whereas there is little spending for this example at interim analysis.

\begin{table}
\begin{center}
\caption{\label{tab:tab1}Comparison of Bonferroni-adjusted vs. CCS-Adjusted Design.}
\begin{tabular}{l rr | rr}
\hline
 Value & \multicolumn{2}{c}{Bonferroni}  & \multicolumn{2}{c}{CCS}  \\
\hline
 &Subgroup &Overall &Subgroup &Overall\\
\hline
 Nominal $\alpha$ level & 1.25\% & 1.25\% & 1.53\% & 1.53\% \\
 Population power\footnote{footnote text} &90\% &90\% &91.44\% &91.36\%\\
 Number of events$^2$ &296&434&283&415\\
\hline
\end{tabular}
\begin{tabular}{l rrr | rrr}
 &IA1 &IA2 &FA &IA1 &IA2 &FAl\\
\hline
Z statistic bound &3.35 &2.67 &2.28 &3.24 & 2.58 &2.21\\
Hazard ratio bound &0.62 &0.73 &0.79 &0.63 &0.74 &0.80\\
\hline
\end{tabular}
\begin{tabular}{l}
\footnotesize
$^1$ With fixed HR(0.65/0.70) and sample size (296/434) in subgroup/overall population.\\
\footnotesize
 $^2$ With fixed HR (0.65/0.70) and power (90\%) in subgroup/overall population.
\end{tabular}
\end{center}
\end{table}

The nominal alpha level increases from 1.25\% to 1.53\% when the CCS is adjusted in the group sequential design and FWER is controlled at the same level of 2.5\%.  The impact of the adjusted nominal alpha level is reflected in Z statistic bounds as well.  For example, the bound at IA1 decreases from 3.35 to 3.24 after adjusting for the CCS.  In Table 1, we present the comparison of hazard ratio bounds showing CCS in group sequential design has a minor release and flexible restriction in hypothesis testing.  In addition, the gain from CCS group sequential design includes a smaller required number of events with a 4.4\%-saving in each population.  Lastly, population power increases from 90\% to 91.4\% with fixed hazard ratio and sample size in both subgroup and overall populations.  

The cost of a clinical trial is always an essential concern, and one way to control the budget is recruiting minimum but sufficient patients in the trial.  The example above shows a 19-event saving in the overall population after applying CCS in group sequential design.  When calculating the number of patients required in this example, a 33-patient (721-688) savings is realized.  A recent report in Journal of Clinical Oncology \cite{Steensma2014} showed the cost of a phase IIIA oncology trial ranged from \$75,000 to \$125,000 per patient.  Assuming the cost per patient in a trial is \$100,000, we would save \$3.3 million with the application of CCS group sequential design in this example. Alternatively, a 1.36\% power increase in Table 1 applied to a trial that could result in a drug approval worth \$100 million would have a value of \$1.36 million.

\section{Extensions}

\subsection{More Complex Graphical Approaches}

When multiple endpoints as well as multiple populations are considered in a trial, we suggest applying CCS to the hypotheses with the same endpoint.  
For instance, Figure 4 shows a hypothetical case using a graphical multiplicity approach similar to Example 2 in \cite{Bretz2011} that includes only part of the correlation matrix known. 
There are two endpoints (OS and progression-free survival (PFS)) and two populations. CCS needs to be conducted twice, one time for OS and another time for PFS.  
For OS, there is a total alpha of 2\% available. Splitting this equally, we can test for OS differences initially using the CCS methods shown above. 
The transfer of nominal alpha once a hypothesis has been rejected is not impacted by CCS; e.g., if H2 were rejected, then H1 may be tested at level alpha=2\%.  
Similarly, if H1 and H2 both are rejected, there is total 2\% alpha level that can be transferred to H3 and H4 equally, e.g. 1\% alpha level to H3, and 1\% alpha level to H4. 
Since there are only 2 populations, there are still no consonance issues. 
The gain of nominal $\alpha$-level from CCS (e.g. CCS-$\alpha$ of 1.1\% in H1 and H2 or CCS-$\alpha$ of 0.3\% in H3 and H4) is only available to the hypothesis testing within the endpoint itself.  

\setcounter{figure}{3}
\begin{figure}
\centering
\caption{Graphical Approach with CCS Adjusted $\alpha$-levels.}
\includegraphics{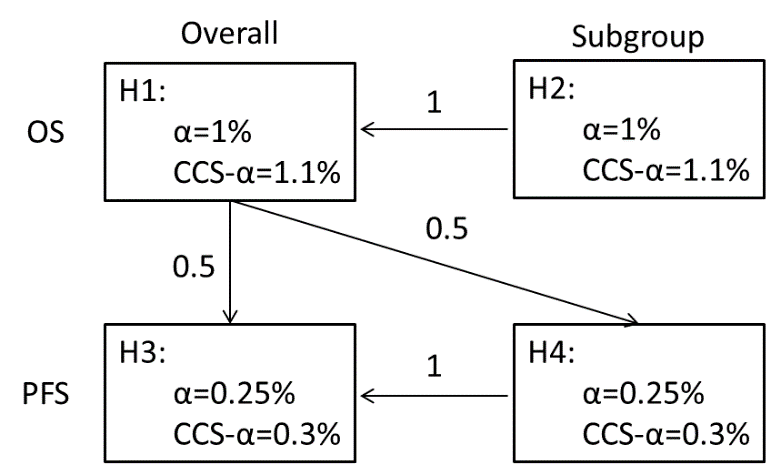}
\end{figure}

\subsection{Multi-Arm Multi-Stage Multi-Population (MAMSMP) Trials}

We focus on 2-arm trials in previous sections, but the weighted parametric closed testing procedure has been extended to multiple-arm designs \cite{Ghosh2017, Wason2016, Jaki2019}.
Based on the results here, multiplicity for multi-arm, multi-stage, multi-population (MAMSMP) designs can fully account for correlations when designing and analyzing the trial.
For example, consider a 3-arm, 2-population trial comparing 2 dose arms (e.g., high-dose and low-dose) against a control arm when one interim analysis is planned.  
Let $Z_{Hik}$ be the test statistics when comparing high-dose with control, and $Z_{Lik}$ be the test statistics for comparing low-dose with control.  The CCS correlation matrix is a $8\times 8$ matrix for the 8 test statistics ({\it i.e.,} 
$(Z_{H11}, Z_{H12}, Z_{H21}, Z_{H22}, Z_{L11}, Z_{L12}, Z_{L21}, Z_{L22}).$  
The covariance of any two of above test statistics is as explained in Section 3.1.  
Particularly, the covariance of $(Z_{Hik}, Z_{Li^\prime k^\prime })$ is the correlation from the control observations that are included in both high-dose and low-dose test statistics.  
Once the CCS matrix is known, calculations described in the Section 3 for group sequential monitoring boundaries, power, and sample size can be applied to MAMSMP design. 

\section{Discussion}

In this paper, we have introduced the complete correlation structure (CCS) to manage both temporal and subpopulation correlations among test statistics in group sequential design.  
To our knowledge, this is the first paper that has accounted for these two aspects simultaneously in a clinical trial design for nested subgroups.
The approach has been used previously for multi-arm multi-stage trials \cite{Ghosh2017, Wason2016, Jaki2019}.   
By synthesizing concepts from the literature, we have built upon the CCS to compute group sequential boundaries, population power, and sample size with weighted parametric closed testing procedure using the graphical approach for multiplicity control \cite{MaurerBretz2013}.  
The advantages of using the CCS include more relaxed efficacy boundaries ({\it i.e.,} greater nominal alpha level), higher population power, or smaller required sample size for a given FWER.  
The method can be applied to multiple types of outcomes (e.g. normal, binary, and survival type) since it is  based on standardized test statistics with the same asymptotic properties.  

We have examined the influence of biomarker prevalence on nominal alpha level, population power, and sample size (Figure 1 – Figure 3).  
The impact of the proportion in the subgroup was more influential than group sequential adjustment over time in interim analysis, probably due to the small amount of interim $\alpha$-spending for the temporal analyses.  
In group sequential design with the Bonferroni-style of FWER control, we account for the correlation among test statistics at different stages, but the correlation between populations is not incorporated.  

When the nominal alpha level is elevated by CCS in group sequential design, the boundary values for hypothesis testing are lowered.  
In the example at Section 5, the nominal alpha level  increases by 0.28\% in each population.  
This improvement increases the chance for a positive efficacy finding.  
Corresponding HR approximations at bounds are also less stringent when FWER is controlled at the same 2.5\% significance level. 
While these differences may appear minor, narrow misses for statistical significance can be extremely costly in terms of lost opportunity for regulatory approval. 
Also, the amount the bounds are relaxed are not so different than what can be obtained with group sequential testing accounting for temperal correlations (standard practice) vs. using a Bonferroni adjustment for group sequential testing. 
The financial savings of \$3.3 million is important and the cost saving would increase if biomarker prevalence is higher than our assumption at 60\%.  
In terms of population power, we have showed CCS in group sequential design had greater power to detect treatment effect while sample size and effect size were fixed, again confirming the cost-effectiveness of this approach.   

We have discussed the extension of CCS in Section 6 that CCS is suitable to multiple-arm design and more complex graphical approach with some constraints.  
These options make application of CCS feasible for a variety of clinical trial designs, a subject of ongoing research.  

Unfortunately, off-the-shelf software does not provide tools to derive designs or testing for trials incorporating CCS for nested subgroups. 
However, the calculations are not terribly complex using readily available tools in R such as a combination of the {\bf gsDesign} package \cite{gsDesign} for group sequential design, the {\bf gMCP} package \cite{gMCP} for graphical hypothesis testing and the {\bf mvtnorm} package \cite{mvtnorm} for multivariate normal probability calculations.

\section{Conclusion}

The complete correlation structure (CCS) approach simultaneously incorporates correlations between populations and interim timing into group sequential design. 
It has an extensive application to the methods that are widely used in current clinical trials, and it is applicable to designs with multiple arms, multiple stages, and multiple populations. 
It can also can be integrated with the graphical testing approach with the qualification that for more than 2 populations a full closed testing evaluation may be required.  
The gains in efficiency of applying the CCS in group sequential design as compared to conventional group sequential design includes more relaxed efficacy boundaries 
({\it i.e.,} larger nominal $\alpha$-level at each test), resulting in greater power or reduced sample size that are small, but meaningful; in fact, the gains from considering population correlations can be greater than those achieved by incorporating broadly-used temporal correlations.

\end{document}